# About accuracy of the solution of NP-complete tasks

Rustem Valeyev, Saint Petersburg

## Subject

From the point of view of complexity of algorithm of discrete tasks exact decision, for which the polynomial of the task's dimensions (polynomial from length of an input) algorithms are existing, creates the class P. Discrete tasks, for which such algorithms are unknown, creates the class NP. Dimension here is understood as certain parameter (or parameters) which anyhow defines or quantity of examined objects, or - quantity of actions which essentially should be presented at algorithm of the decision of a task. More particularly for the general case here nothing is possible to say, because virtue of that the variety of the tasks which are included in classes P and NP, is rather great.

Another formulations are constructed in terms of Turing machine: P is a class of tasks, which have polynomial solving by determined Turing machine, and NP - a class of tasks, which have polynomial solving by not determined Turing machine.

In the class NP, as a subset, includes NP-complete tasks, to which polynomially reduce any tasks from NP. NP-complete tasks, with the full-proofed substantiation, are considering the hardest soluted tasks in NP class.

For some discrete task it is not possible to design polynomial (effective) algorithm despite of serious efforts of many mathematicians. What is the reason of these failures? Such algorithm does not exist in nature, or polynomial (effective) algorithm exists but find him is very difficultly? This found fundamental problem got a name «problem P versus NP» [1] ... [4]. Inaccessibility of this riddle, numerous works in the field of development of approximate algorithms and absence of effective algorithms for class NP leave other circle of questions which are very important for tasks on discrete structures in a shadow. Namely: that represent objectively existing exact and approached answers of these tasks and that it is possible to expect from the algorithms intended for their detection.

## 1. Nonstrict definition of tasks of class NP

Widely used informal formulation concerning the problem «P vs NP» and to the questions connected to it, looks like the following: P-tasks are discrete tasks, which have answer and it is easy to find (exists polynomial algorithm of reception of the answer) and at which validity of this answer is easy for check up. NP-tasks are discrete tasks in which validity of the answer it is easy to check up (exists the effective algorithm of check of correctness of this answer), but this answer was found by unknown way in an available kind.

## 2. Necessary terms

For brevity, accuracy and unambiguity the following working terms further are used.

***accessibility of parameter A to the developer of algorithm*** - opportunity for developer of a mass task algorithm to realize application for all possible values of parameter A in order to get an exact answer for any of the individual cases of this mass task.

***accessibility of set B to the developer of algorithm*** - accessibility of all parameters of all members of set B.

Example of accessible and inaccessible sets (sets of the data): mathematical model of motion of every ball playing billiards - and the same for every molecule in the Brownian motion in one drop of water.

***N-dimensional space of a task*** - Cartesian product of N sets; each of these sets is the set of all the values of one of the N (where $N \geq n$) variables that exist in the task (n is the quantity of unknown); synonym: «set of possible answers of a task», «senior space».

***possible answer [of a task]*** - any point of N-dimensional space of this task.

***forbidden answer [of a task]*** - possible answers of the task which can not be its exact answer.

***allowable answer [of a task]*** - possible answer of the task which is not a forbidden answer.

***exact answer [of a task]*** - a subset of set of allowable answers; each set member of this subset completely satisfies to all those requirements to exact answer which contain in initial data of this task.

***exhaustive method -*** detection of any and all set members of the set and the determination of value (values) of the parameter (parameters) of each separate member of the set.

***exhaustive search –*** algorithm which uses the exhaustive method.

Even the most successful and respected algorithm can't be regarded as exact (i.e. not only as effective, but as full too) if it can not guarantee reception of the exact answer of everyone without exception of an individual case of a considered general problem. In mathematics by default by the task's answer is meant its exact answer, and as algorithm of the solution of a task – its exact algorithm.

By consideration of basic possibility of the solution of tasks on discrete structures the exponential algorithm "exhaustive search" (which is inefficient exact algorithm) to some extent plays a role of the not determined Turing machine (and not determined Turing machine - a role of an all-powerful all-knowing magic wand) – and all of them together or/and separately represent a certain help from the oracle.

As evident illustrations five tasks which have essentially different internal structure from the point of view of the accuracy of their decision are mentioned further. All algorithms intended for their decision are meant technically perfect, i.e. not containing basic and technical mistakes and capable of obtaining exact answers as far as the best algorithms of such type are theoretically capable of it.

Most often questions of accuracy concern these or those problems relating to behavior of allowable answers in the next vicinities of the exact answer, and also to compliance of objective reality of those declared functional dependences which define these allowable answers (Fig.1 and 2). But sometimes a "the next vicinities" in which in an unpredictable way exact answers take place, are all exponential set of possible answers of a task. Parameters of members of such set, allowing to identify the exact answer, don't submit to one general for all values of this parameter to the rule [10]. For example, in many tasks of the class P and the class NP (see Fig.4 and 6 respectively).

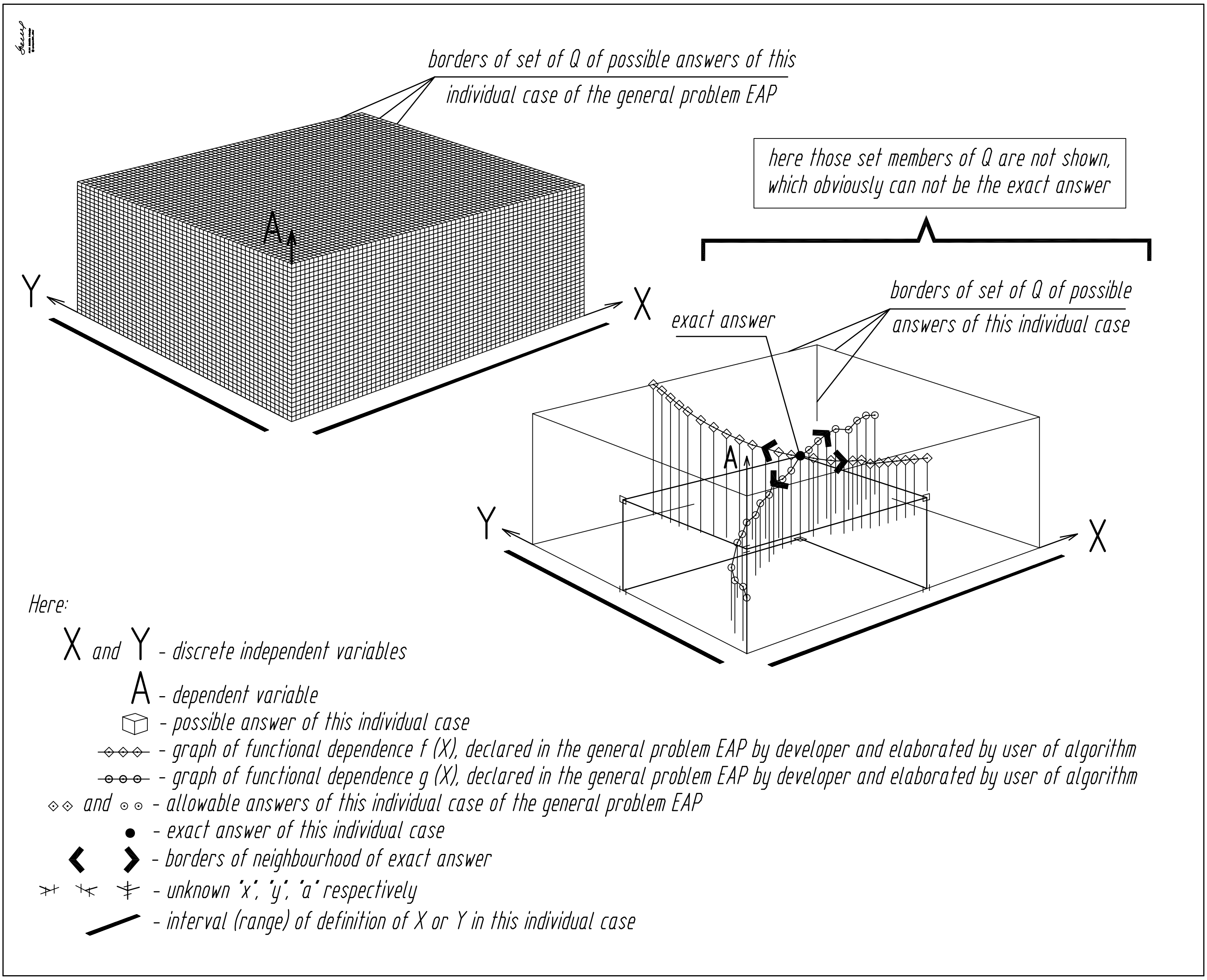

*Fig.1 All variables, set Q of possible answers (i.e. N-dimensional space, N=3, n=2) and neighbourhood of the exact answer in an individual case of the general problem EAP*

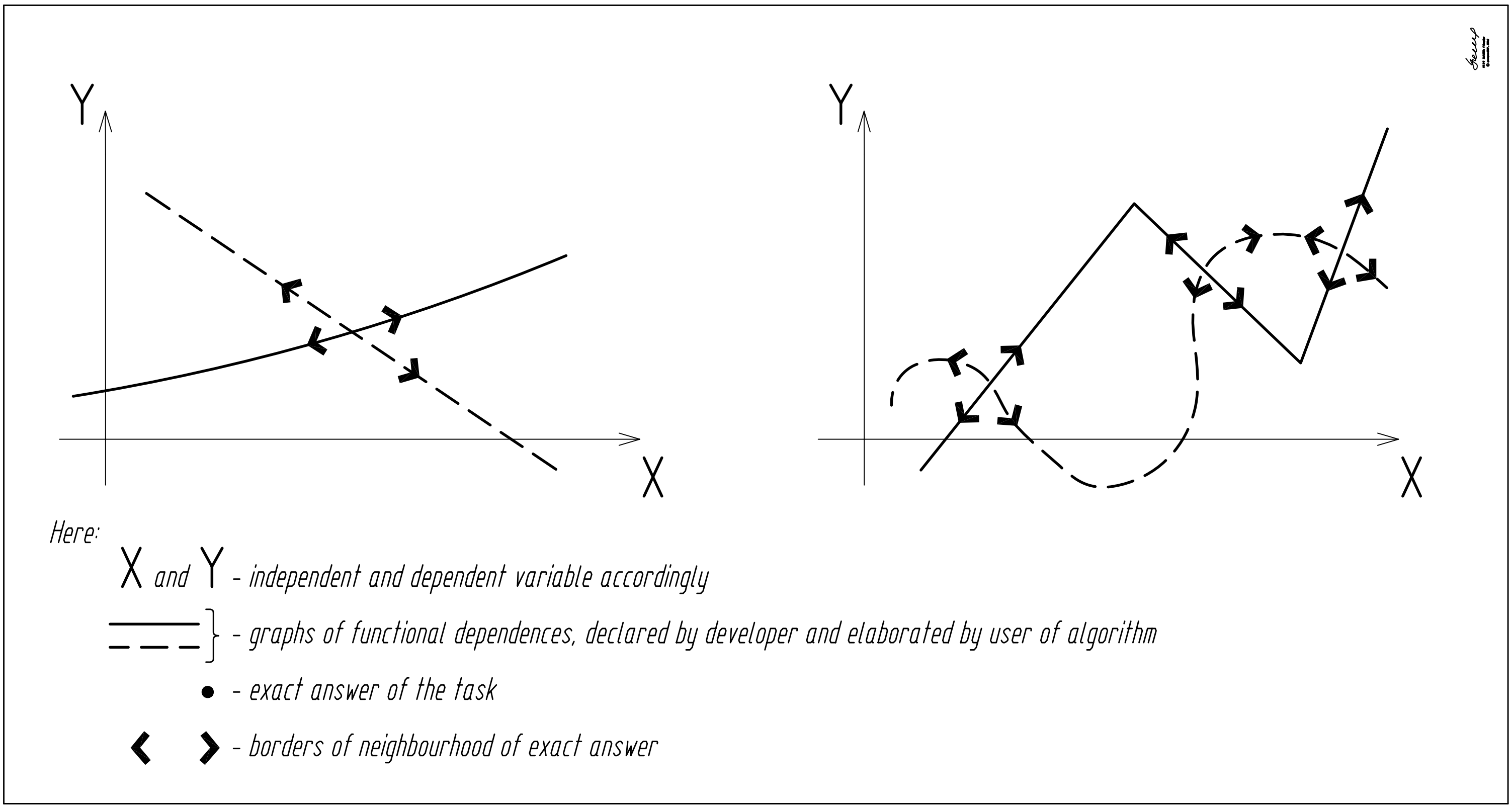

*Fig.2 An examples of neighbourhood of exact answer in some mathematical tasks (N=2, n=1)*

The concept «accuracy of the answer of a task» supposes many various ways of formalization. It is possible to use, for example, the following approach:

***radial accuracy [of an allowable answer] -*** the quantitative parameter of the allowable answer equal to an absolute value of a difference between values of the same $P_{ea}$ parameter of the allowable answer, received by algorithm, and the exact answer, where $P_{ea}$ is the parameter by means of which the exact answer is defined.

When in a task exists more than one allowable answer, all members of this set can be naturally presented in the form of the certain ordered sequences. It is most convenient to have allowable answers according to change of value of parameter by means of which among allowable answers the exact answer can be identified. To each number (position) in such sequences more than one allowable answer can be put in compliance: if they possess the same value of the specified parameter.

It is fair for many discrete problems with additional quantitative requirements to the exact answer: for class P (task SRP), for NP-complete tasks (TSP, "Knapsack Problem", "Partition" etc) etc.

Therefore in such tasks in addition to usual quantitative or/and analytical estimates of accuracy of the answer received by this or that algorithm (i.e. to estimates of discrepancy of values of parameters of the exact and received answers) it is possible to use the parameters similar to estimates of sports and those similar achievements, to various ratings, lists, etc.

***rating position [of an allowable answer] -*** position of the allowable answer in a sequence ordered on increase or on decrease of value of that parameter which defines the exact answer of a task.

***positional accuracy [of a allowable answer] -*** rating position of the allowable answer provided that to the exact answer corresponds rating position 1.

As natural development of concepts «radial accuracy» and «positional accuracy» it is possible to use more universal relative parameters «***relative radial accuracy***» and ***«deviation accuracy*** ».

Practical value of all these concepts (as well as concept «radial accuracy») is insignificant, but they are very convenient by theoretical consideration of the questions connected to accuracy of the solution of tasks on discrete structures. Especially when there is an opportunity to compare beforehand known exact answer with that result to which the considered algorithm has come.

## 3. Accuracy of the answer of a classical algebraic task

***A typical algebraic task («Elemental Algebraic Problem», EAP)***, see Fig.1: two or more variable are set and certain requirements to the required answer, expressed in the form of functional dependences between these independent and dependent variables and other conditions. By means of these or those mathematical manipulations and calculations it is necessary to reveal crossings of those sets which values of the mentioned functions are.

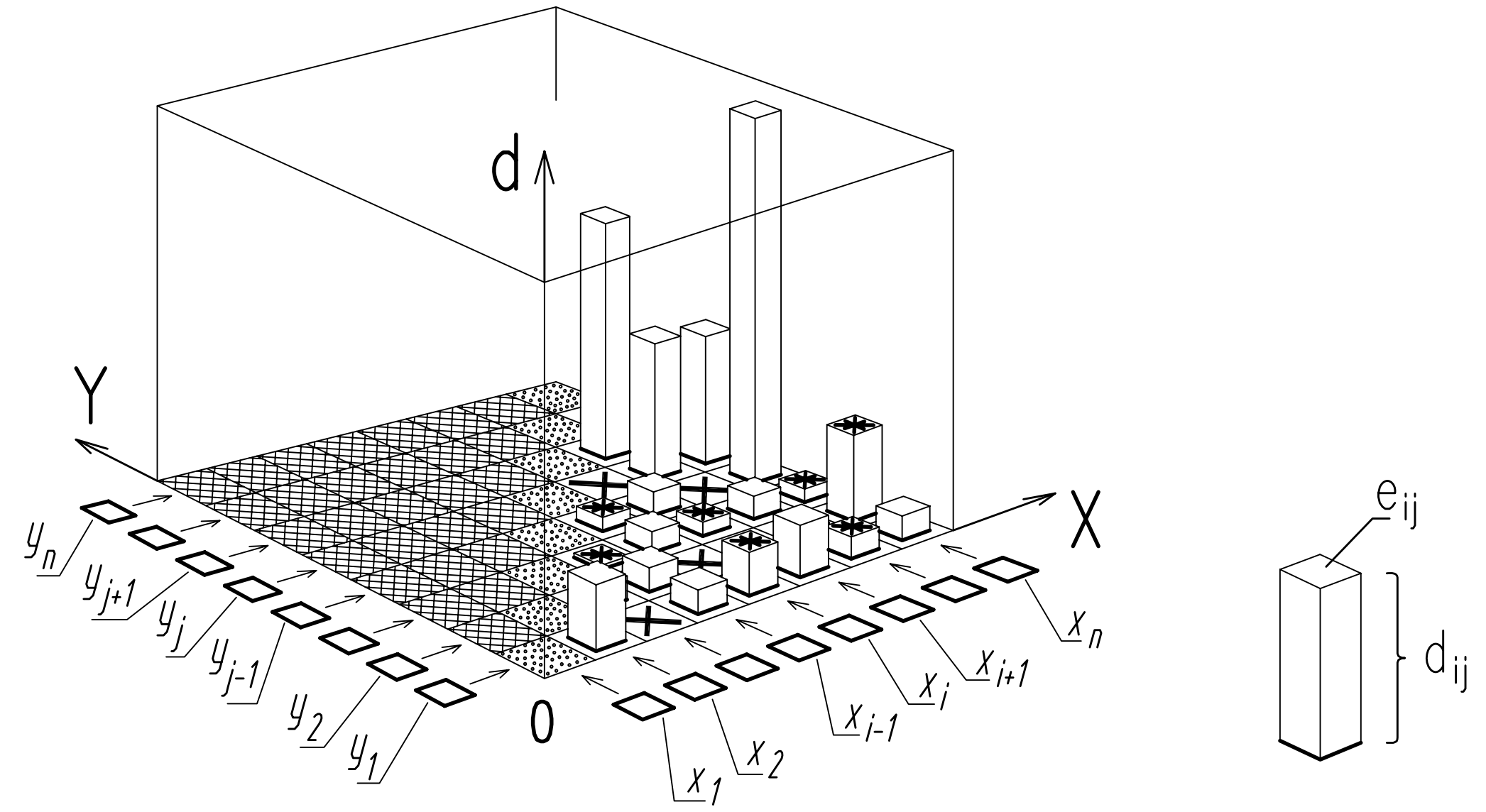

Here:

- $x_i$ - ij-th set member of set "independent variable X"; parameter $P_1$ appropriated to this set member: "i-th initial graph node"
- $y_j$ - j-th set member of set "independent variable Y"; parameter $P_1$ appropriated to this set member: "j-th final graph node"
- $d_{ij}$ - ij-th set member of set "independent variable d"; parameter $P_1$ appropriated to this set member: "length of a tree edge between i-th initial and j-th final graph nodes"
- $e_{ij}$ - ij-th set member of set E "unforbidden tree edges"; user's parameters are convenient for interpreting as vectors with the beginning in the point 0
- - ij-th set member of set T "forbidden tree edges"
- - set member of set UE "nonexistent tree edges"
- - set member of set UL "not making sense tree edges" (edge which have no useful physical sense for algorithm)
- - tree edges which in the given individual case of general problem SRP consistently are chosen by greedy algorithm

Fig.3 The set of independent variables in the task SRP (interpretation as the three-dimensional ordered space)

Note: in this figure those possible answers of the task are not show, which knowingly can not be the exact answer

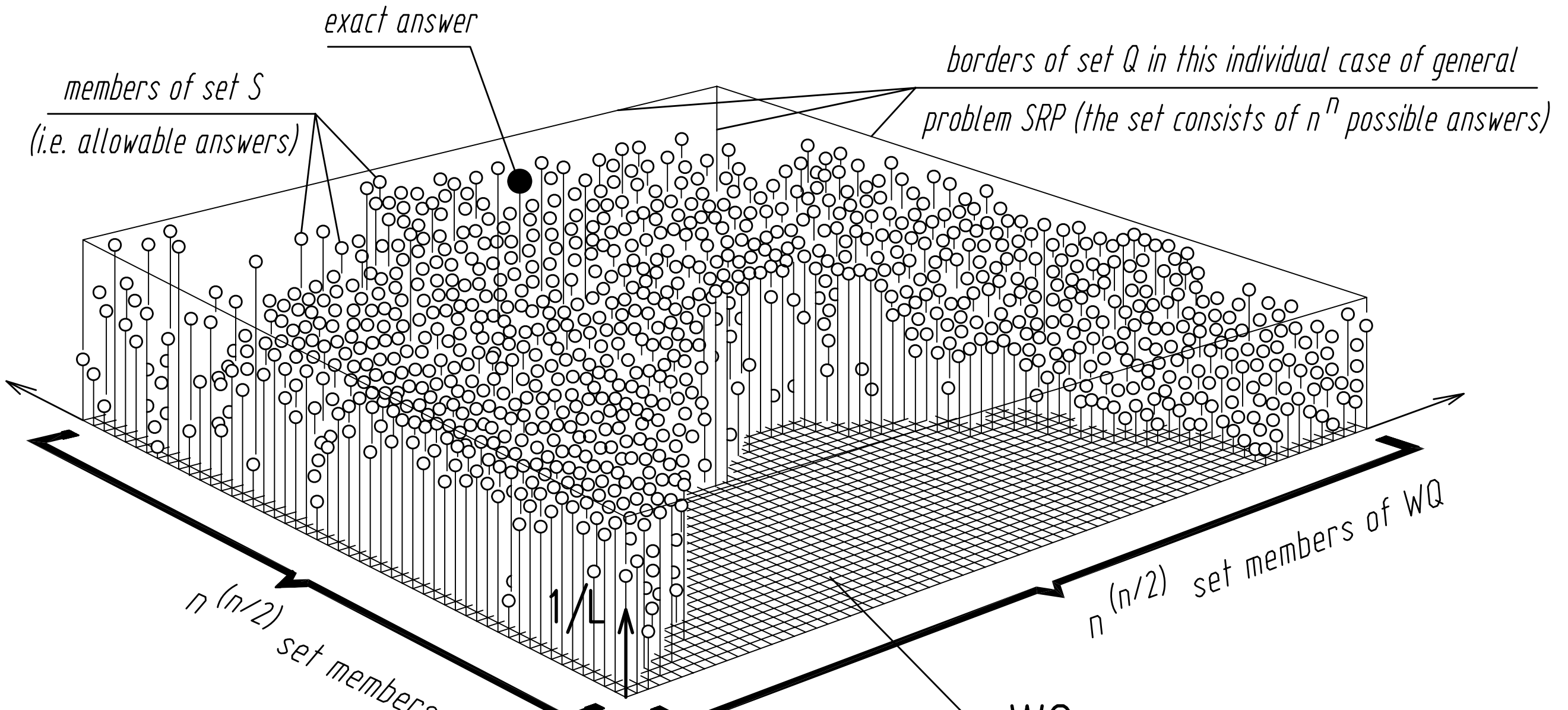

Here:

- Q - set of possible answers
- S - set of allowable answers
- WQ - ordered set at which each set member is unique formally possible sequence "anyone n, including repeating, graph nodes"
- L - parameter which defines the exact answer (length of a route)
- ○ - values of parameter L for allowable answers which are not required (i.e. are not the exact answer)
- ● - value of parameter L for that from allowable answers which is the required exact answer

Fig.4 The set of possible answers and neighbourhood of the exact answer in the SRP (in form of the three-dimensional ordered space)

Here:

Allowable answers: exists.

Exact answer (which by definition is a certain subset of allowable answers which completely meet all those requirements to the exact answer which are in the initial data of a considered individual case of the general problem): exists.

The fact of existence of the exact answer of a considered individual case of the general problem: is beforehand known (even before reception of task's unknowns, i.e. before reception of coordinates of crossing of graphs of the declared functions, it is possible to find out, whether these graphs are crossed).

Exhaustive search: always receives the exact answer (and also all allowable answers).

Exact polynomial (effective, on the basis of operations with functional dependences) algorithm: exists**.**

Exact exponential (inefficient) algorithm which is not being exhaustive search: it is not necessary.

Approached polynomial (effective) algorithm: it is not necessary.

Approached exponential (inefficient) algorithm which is not exhaustive search: it is not necessary.

Availability of the exact answer of a task prior to the beginning of work of algorithm (i.e. knowledge of the parameters completely determining the exact answer): parameters of the exact answer are unknown.

Availability of the exact answer of a task after the ending of work of algorithm: parameters of the exact answer are known.

The accuracy got by the best algorithm of the solution of the task EAP (i.e. differences of parameters of the answer received by algorithm from the appropriate parameters of the exact answer): differences do not present (the received answer always represents just the exact answer).

## 4. Accuracy of the answer of a task of class P

The task «Shortest Route Problem» (SRP), see Fig.3, 4: conditions of this task of class P differ from conditions of the NP-complete task TSP ("Traveling Salesman Problem", see below) only that the shortest way shouldn't be closed (i.e. it shouldn't be continued from final graph node back to the initial node).

Here:

Allowable answers: exists.

Exact answer: exists.

The fact of existence of the exact answer of a considered individual case of the general problem: is beforehand known (if each node of the graph possesses at least 2 edges, in such graph there is at least one not closed way; at identical number of edges the way consisting of the smallest edges always has the smallest length).

Exhaustive search: always receives the exact answer (and also all allowable answers).

Exact polynomial (effective) algorithm: exists (for example, the "greedy" algorithm).

Exact exponential (inefficient) algorithm which is not being exhaustive search: it is not necessary.

Approached polynomial (effective) algorithm: it is not necessary.

Approached exponential (inefficient) algorithm which is not exhaustive search: it is not necessary.

Availability of the exact answer of a task prior to the beginning of work of algorithm (i.e. knowledge of the parameters completely determining the exact answer): parameters of the exact answer are unknown.

Availability of the exact answer of a task after the ending of work of algorithm: parameters are known.

The accuracy got by the best algorithm of the solution of the task SRP (i.e. differences of parameters of the answer received by algorithm from the appropriate parameters of the exact answer): differences do not present (the received answer always represents just the exact answer).

# 5. Accuracy of the answer of NP-complete tasks

## 5.1. NP-complete task without additional requirements to the exact answer

***The task «Boolean satisfiability problem»***, ***(«SATISFIABILITY»***, ***SAT)*** - the Boolean formula is given, i.e. finite set of variables, brackets and operations "AND", "OR", "NOT". It is required to find such set of values of variables (values can be only "YES" or "NO"), on which this formula accepts TRUTH value. According to Cook's theorem (it is proved by Stephen Cook, 1971), the problem SAT for the Boolean formulas which have been written down in a conjunctive normal form, is NP-complete (further SAT is meant as this the first of the found NP-complete tasks).

Here:

Allowable answers: do not exists.

Exact answer: can exist and can not exist.

The fact of existence of the exact answer of a considered individual case of a general problem: in general case is not known beforehand.

Exhaustive search: always receives the exact answer if it exist (and also all allowable answers).

Exact polynomial (effective) algorithm: now it is not known.

Exact exponential (inefficient) algorithm which is not exhaustive search: now it is not known (or it is poorly known).

Approached polynomial (effective) algorithm: it is impossible.

Approached exponential (inefficient) algorithm which is not being exhaustive search: it is not necessary.

In SAT with its extreme ascetic features and nonflexible "black-and-white" logic any allowable answer is exact answer. Therefore such as SAT "Boolean" tasks without additional conditions is so difficult to solve: additional conditions, whatever they are, generate an exponential subset of those allowable answers which don't exact (i.e. a subset of approximate answers).

Availability of the exact answer of a task prior to the beginning of work of algorithm (i.e. knowledge of the parameters completely determining the exact answer): parameters of the exact answer are unknown.

Availability of the exact answer of the task after the work of algorithm ending:

if it is proved, that «P = NP», than parameters of the exact answer are known;

if it is proved, that «P $\neq$ NP», than in general case the task SAT can't be solved

The accuracy got by the best algorithm of the solution of the task SAT (i.e. differences of parameters of the answer received by algorithm from the appropriate parameters of the exact answer):

if it is proved, that «P = NP», than differences do not present (the received answer always represents just the exact answer);

if it is proved, that «P $\neq$ NP», than in general case the task SAT can't be solved.

## 5.2. NP-complete task with additional quantitative requirements to the exact answer

***The task "Traveling Salesman Problem" (TSP),*** see Fig.5, 6: the graph with n node and with arbitrary number of arbitrarily located edges of any length is set. It is necessary to find such sequence from n of not repeating edges ("route") which begins in a certain initial node, consistently visits all other nodes and comes back again to initial node. It passes only one time through each node and at the same time has the smallest length from all other such routes, possible on this graph.

Here:

Allowable answers: exists.

Exact answer: exists.

The criterion on which among set of allowable answers it is possible to identify the exact answer, is objective and does not depend on desire of the developer or the user of algorithm: it is condition «the route should have the least length among other routes in the graph).

The fact of existence of the exact answer of a considered individual case of a general problem: is beforehand known (if each graph node has at least 2 edges, than in such graph there is at least one closed route; if there are some such routes at least one of them has the least length).

Exhaustive search: always receives the exact answer (and also all allowable answers).

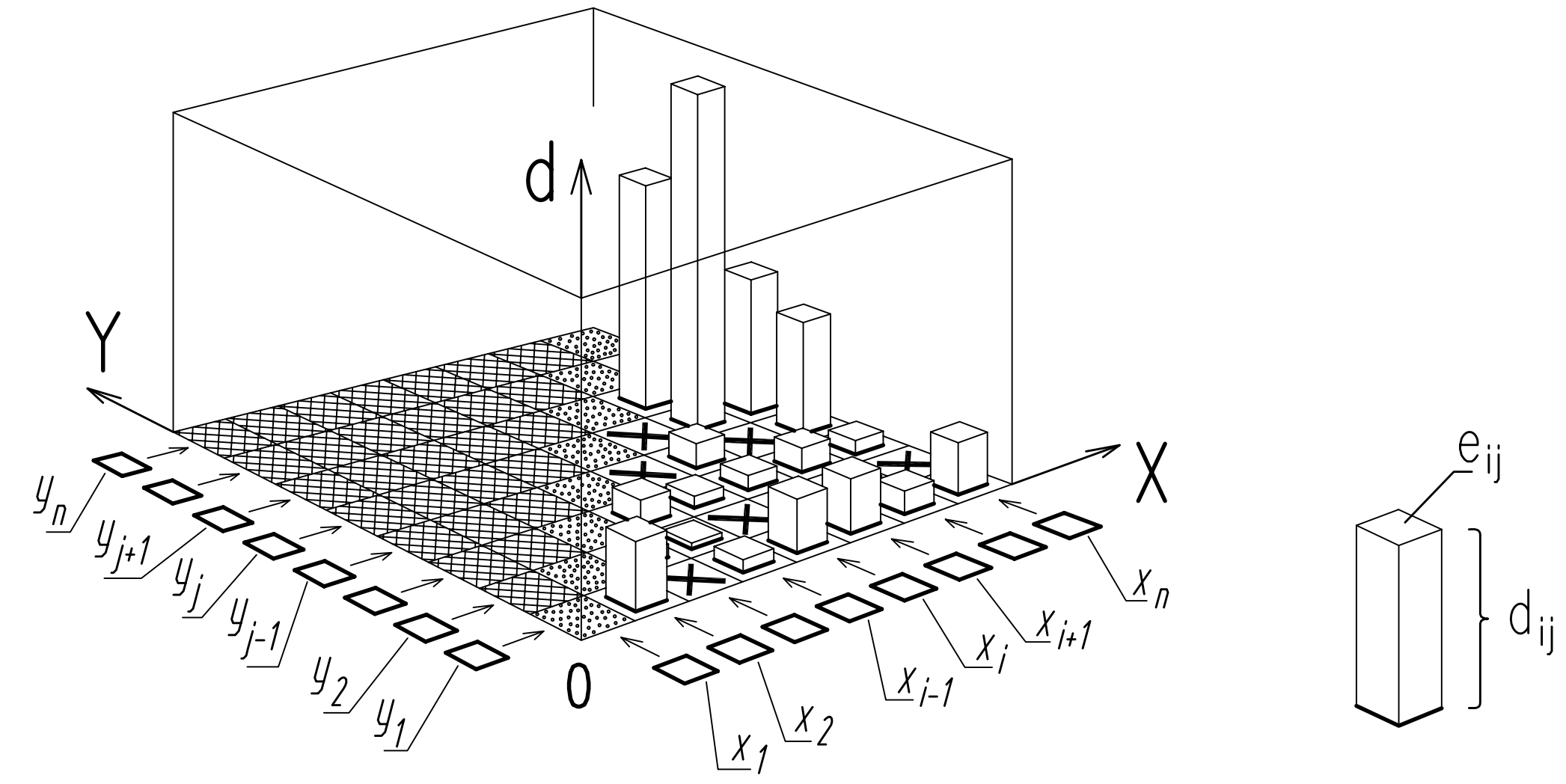

*Here:*

$x_i$ – *ij-th set member of set "independent variable X"; parameter $P_1$ appropriated to this set member: "i-th initial graph node"*

$y_j$ – *j-th set member of set "independent variable Y"; parameter $P_1$ appropriated to this set member: "j-th final graph node"*

$d_{ij}$ – *ij-th set member of set "independent variable d"; parameter $P_1$ appropriated to this set member: "length of a tree edge between i-th initial and j-th final graph nodes"*

$e_{ij}$ – *ij-th set member of set E "unforbidden tree edges"; user's parameters are convenient for interpreting as vectors with the beginning in the point 0*

– *ij-th set member of set T "forbidden tree edges"*

– *set member of set UE "nonexistent tree edges"*

– *set member of set UL "not making sense tree edges" (edge which have no useful physical sense for algorithm)*

*Fig.5 The set of independent variables in the task TSP (interpretation as the three-dimensional ordered space)*

*Note: in this figure those possible answers of the task are not show, which knowingly can not be the exact answer*

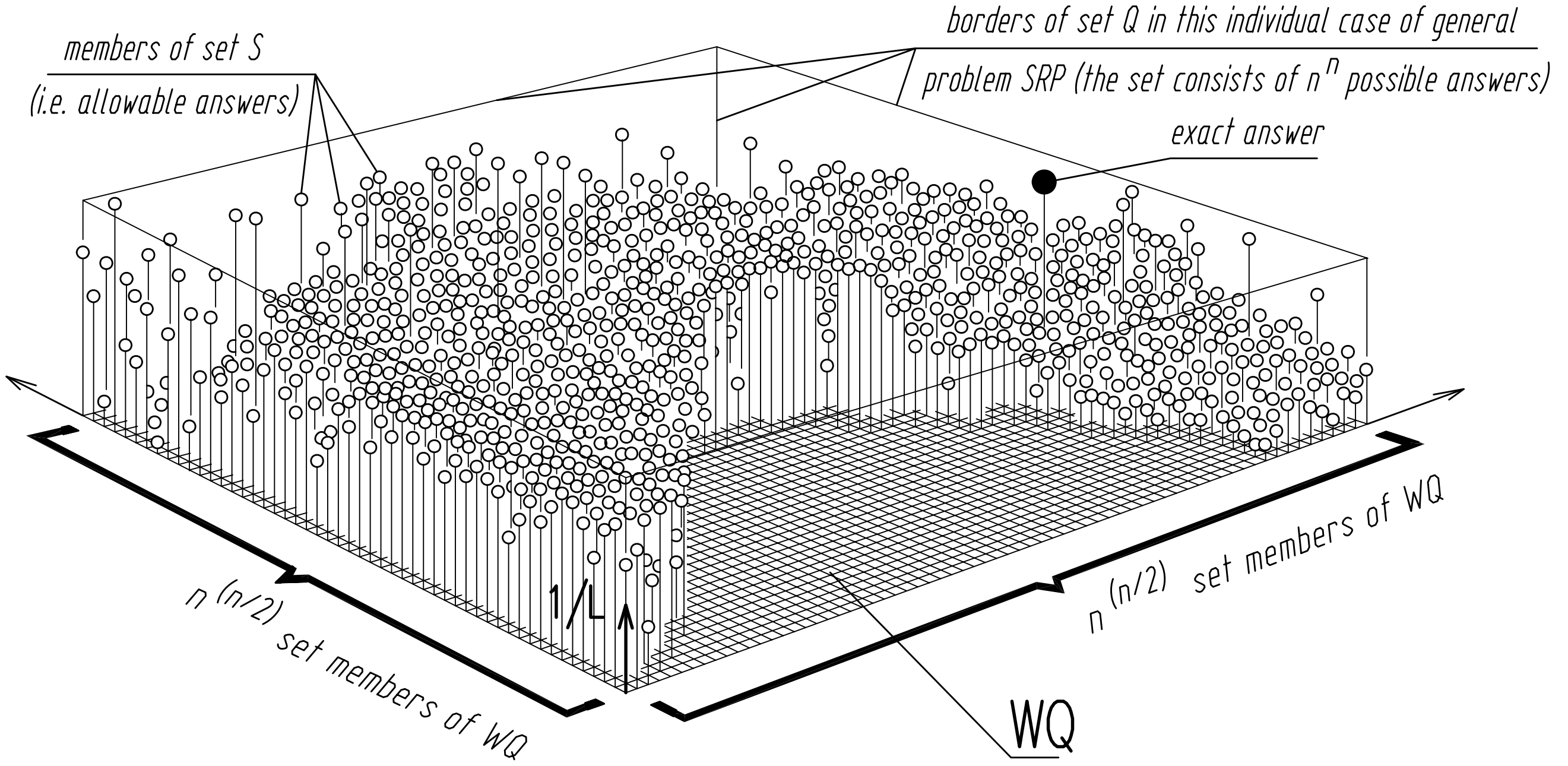

*Here:*

Q – *set of possible answers*

S – *set of allowable answers*

WQ – *ordered set at which each set member is unique formally possible sequence "anyone n, including repeating, graph nodes"*

L – *parameter which determines the exact answer (length of a route)*

○ – *values of parameter L for allowable answers which are not required (i.e. are not the exact answer)*

● – *value of parameter L for that from allowable answers which is the required exact answer*

*Fig.6 The set of possible answers and neighbourhood of the exact answer in the TSP (in form of the three-dimensional ordered space)*

Exact polynomial (effective) algorithm: just now it is not known.

Exact exponential (inefficient) algorithm which is not exhaustive search: now it is not known.

Approached polynomial (effective) algorithm: it is known.

Approached exponential (inefficient) algorithm which is not being exhaustive search: it is known.

Availability of the exact answer of a task prior to the beginning of work of algorithm (i.e. knowledge of the parameters completely determining the exact answer): parameters are unknown.

Availability of the exact answer of the task after the ending of work of algorithm:

if it is proved, that «P = NP», than parameters are known (the received answer always represents just the exact answer);

if it is proved, that «P $\neq$ NP», than in questions concerning the exact answer of a problem of TSP (and many others "not-Boolean" tasks with additional conditions) arise curious paradox.

Mentioned popular informal nonstrict definition "the class P is the discrete tasks which answer it is easy to find and it is easy to check justice of this answer, and the class NP is the discrete tasks, which justice of the answer (if this answer is somehow already received) easy to check, but it is difficult to find this answer", true only partly.

Even without having the reliable answer to a dilemma «P ? NP» it is simple to notice that such formulation is fair only for some of those tasks of the class NP which mean the Boolean answer "yes/no". And when it "yes/no" it appears quite enough. Example: "Boolean" task without additional requirements SAT.

For the vast majority of tasks of the class NP at which the fact of accessory of their some allowable answers to a subset "the exact answer" depends on quantities, check is physically impossible. Anyway until fast ways will be found (i.e. effective) to solve such problems. In "not-Boolean" NP-complete task TSP each route (i.e. each allowable answer of this task) generally has unique length. Therefore shown (a lot, Turing machine, the oracle, etc) a route can't be identified as the exact answer if it is foreknown length of the shortest route – i.e. for such identification this task has already solved precisely.

Example: for a certain individual case of TSP (initial data are given in Fig.7) the route (ring) with length L=1 166 122 and with sequence of nodes "1-9-6-7-3-8-5-10-2-4-12-11" is shown. The biggest that the owner of this gift of the oracle can make is to be convinced that and really will close this route, includes n=12 of not repeating nodes, consists of n of the edges declared in initial data, and that its length and really is equal 1 166 122. From the point of view of the above-mentioned popular nonstrict informal formulation this problem of the class NP is successfully solved. But, unfortunately, the gift of the oracle appeared the worst of all possible options (from all allowable answers): rating position of this route is equal 6237, it is one of 24 longest among 149 688 routes (the longest among 6237 rings), existing in this individual case of TSP. Exact answer (rating position 1): L = 1 158 524, "1-5-2-4-7-8-11-10-12-9-3-6".

Concrete data (including the exact answer) for this individual case are given in Fig.7. In Fig.8 similar information for another individual case of the general problem TSP is provided. Differences in their initial

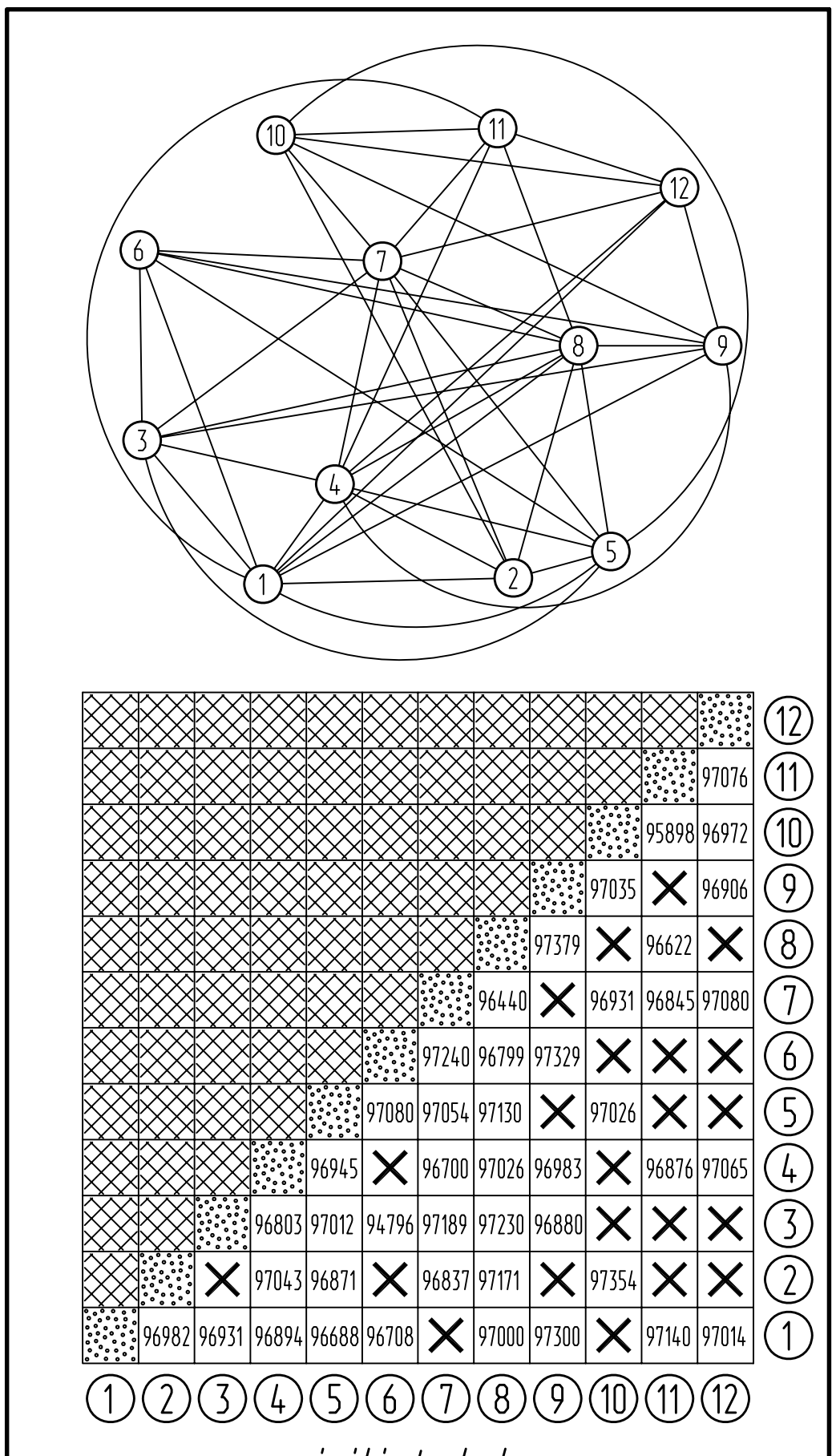

initial data

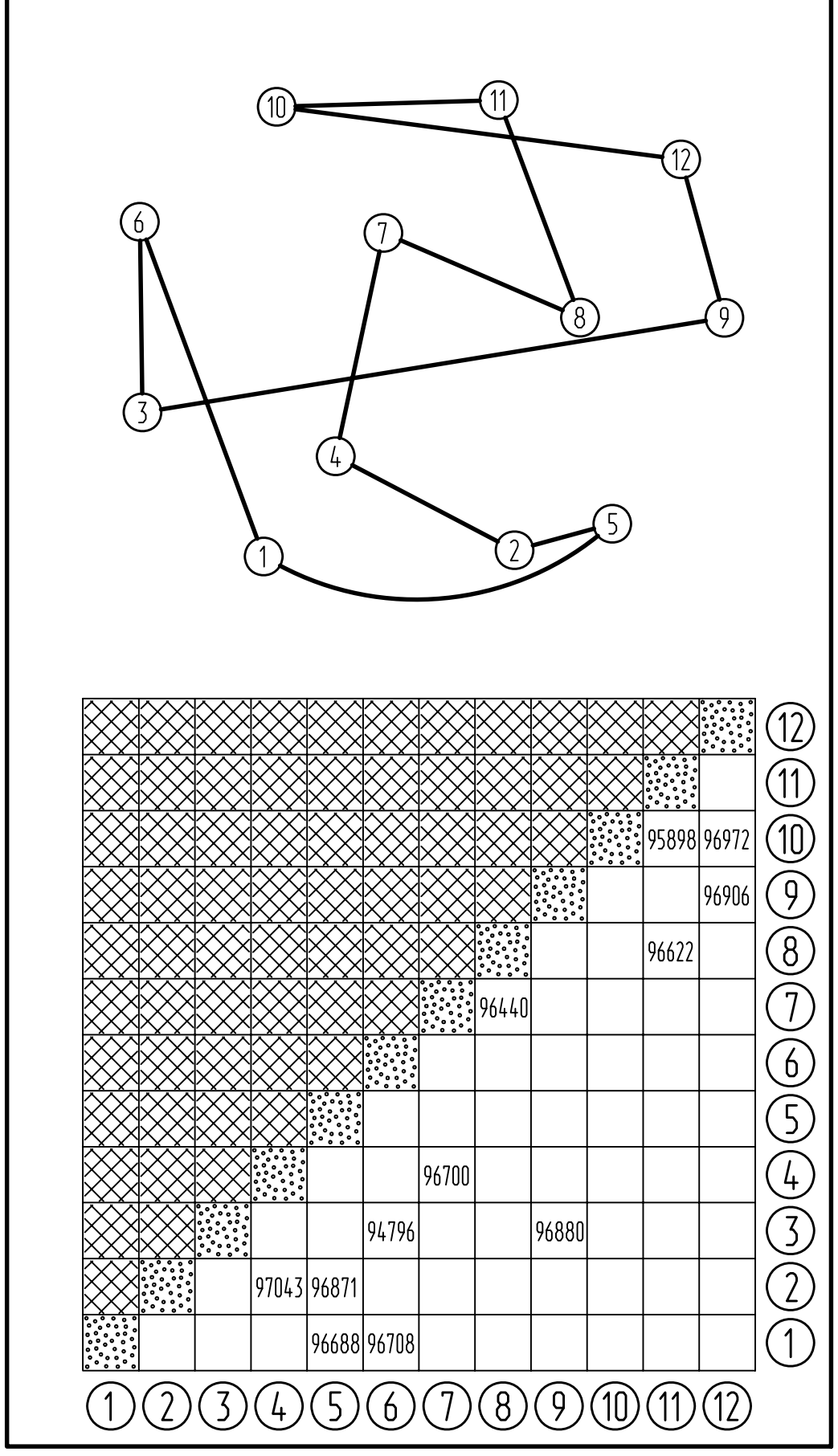

exact answer

## all allowable rings

| r. pos. of L | sequence of graph nodes | L |
|---|---|---|
| 1 | 1-5-2-4-7-8-11-10-12-9-3-6 | 1 158 524 |
| 2 | 1-5-2-7-8-11-10-9-12-4-3-6 | 1 158 669 |
| 3 | 1-2-5-4-7-8-11-10-12-9-3-6 | 1 158 720 |
| 4 | 1-5-4-2-7-8-11-10-12-9-3-6 | 1 158 735 |
| 5 | 1-5-2-7-8-11-10-12-4-9-3-6 | 1 158 760 |
| 6 | 1-4-5-2-7-8-11-10-12-9-3-6 | 1 158 769 |
| 7 | 1-4-11-10-12-9-3-6-8-7-2-5 | 1 158 857 |
| 8 | 1-6-3-4-5-2-7-8-11-10-9-12 | 1 158 875 |
| 9 | 1-2-5-10-11-8-7-4-12-9-3-6 | 1 158 894 |
| 10 | 1-5-2-7-8-4-11-10-12-9-3-6 | 1 158 898 |
| 11 | 1-5-2-7-4-8-11-10-12-9-3-6 | 1 158 904 |
| 12 | 1-5-2-8-7-4-11-10-12-9-3-6 | 1 158 906 |
| 13 | 1-5-10-11-8-7-2-4-12-9-3-6 | 1 158 909 |
| 14 | 1-5-2-10-11-8-7-4-12-9-3-6 | 1 158 928 |
| 15 | 1-4-9-12-10-11-8-7-2-5-3-6 | 1 158 939 |
| 16 | 1-5-2-7-8-6-3-4-11-10-9-12 | 1 158 963 |
| 17 | 1-6-3-9-4-5-2-7-8-11-10-12 | 1 158 966 |
| 18 | 1-5-2-7-4-3-6-8-11-10-9-12 | 1 158 969 |
| 19 | 1-3-6-8-11-10-12-9-4-7-2-5 | 1 159 003 |
| 20 | 1-2-4-7-8-11-10-12-9-3-6-5 | 1 159 007 |
| 21 | 1-2-7-8-6-3-9-12-10-11-4-5 | 1 159 019 |
| 22 | 1-2-5-7-8-11-10-12-9-4-3-6 | 1 159 035 |
| 23 | 1-2-5-6-3-9-12-10-11-8-7-4 | 1 159 041 |
| 24 | 1-6-3-5-2-4-7-8-11-10-9-12 | 1 159 045 |
| 25 | 1-5-2-7-8-6-3-9-4-11-10-12 | 1 159 054 |
| 26 | 1-4-2-7-8-11-10-12-9-3-6-5 | 1 159 056 |
| 27 | 1-5-2-7-4-9-3-6-8-11-10-12 | 1 159 060 |
| 28 | 1-5-2-4-8-7-11-10-12-9-3-6 | 1 159 073 |
| 29 | 1-5-2-8-7-11-10-12-9-4-3-6 | 1 159 081 |
| 30 | 1-4-2-5-7-8-11-10-12-9-3-6 | 1 159 084 |
| 31 | 1-6-3-4-5-2-7-8-11-10-12-9 | 1 159 098 |
| 32 | 1-2-7-8-11-10-12-9-4-5-3-6 | 1 159 101 |
| 33 | 1-2-7-8-11-10-5-4-12-9-3-6 | 1 159 105 |
| 34 | 1-6-3-9-10-11-8-7-2-5-4-12 | 1 159 111 |
| 35 | 1-2-5-10-11-8-7-12-9-4-3-6 | 1 159 115 |
| 36 | 1-5-2-4-9-12-10-11-8-7-3-6 | 1 159 116 |
| 37 | 1-5-4-7-2-8-11-10-12-9-3-6 | 1 159 123 |
| 38 | 1-5-2-4-12-7-8-11-10-9-3-6 | 1 159 126 |
| 39 | 1-6-3-9-12-10-11-4-5-2-7-8 | 1 159 129 |
| 40 | 1-5-2-4-7-8-6-3-9-12-10-11 | 1 159 133 |
| 41 | 1-5-2-4-3-6-8-7-11-10-9-12 | 1 159 138 |
| 42 | 1-5-2-4-11-10-7-8-6-3-9-12 | 1 159 142 |
| 43 | 1-2-5-6-3-4-7-8-11-10-9-12 | 1 159 147 |
| 44 | 1-3-6-8-11-10-9-12-4-7-2-5 | 1 159 148 |
| 45 | 1-5-2-10-11-8-7-12-9-4-3-6 | 1 159 149 |
| 46 | 1-2-7-8-11-10-9-12-4-3-6-5 | 1 159 152 |
| 47 | 1-2-5-8-7-4-11-10-12-9-3-6 | 1 159 159 |
| 48 | 1-5-6-3-4-2-7-8-11-10-9-12 | 1 159 162 |
| 49 | 1-4-2-5-10-11-8-7-12-9-3-6 | 1 159 164 |
| 50 | 1-3-6-8-7-11-10-12-9-4-2-5 | 1 159 172 |
| 51 | 1-5-2-7-8-9-12-10-11-4-3-6 | 1 159 174 |
| 52 | 1-2-5-7-8-11-10-9-12-4-3-6 | 1 159 180 |
| 53 | 1-5-2-7-8-6-3-4-11-10-12-9 | 1 159 186 |
| 54 | 1-4-12-9-3-6-8-11-10-7-2-5 | 1 159 187 |
| 55 | 1-2-5-10-11-4-7-8-6-3-9-12 | 1 159 188 |
| 56 | 1-6-3-4-2-5-7-8-11-10-9-12 | 1 159 190 |
| 57 | 1-5-2-7-4-3-6-8-11-10-12-9 | 1 159 192 |
| 58 | 1-2-7-8-6-3-9-12-4-11-10-5 | 1 159 193 |
| 59 | 1-2-7-4-9-12-10-11-8-6-3-5 | 1 159 195 |
| 60 | 1-3-6-9-12-10-11-8-7-4-2-5 | 1 159 196 |
| 61 | 1-2-7-4-12-9-3-6-8-11-10-5 | 1 159 199 |
| 62 | 1-5-10-11-4-2-7-8-6-3-9-12 | 1 159 203 |
| 63 | 1-5-2-7-4-9-12-10-11-8-3-6 | 1 159 211 |
| 64 | 1-4-7-2-5-8-11-10-12-9-3-6 | 1 159 214 |
| | | |
| 6224 | 1-4-2-10-5-6-9-8-3-7-12-11 | 1 165 820 |
| 6225 | 1-9-8-2-10-5-6-7-3-4-12-11 | 1 165 823 |
| 6226 | 1-9-10-2-8-3-7-6-5-4-12-11 | 1 165 825 |
| 6227 | 1-9-6-5-10-2-8-3-7-12-4-11 | 1 165 840 |
| 6228 | 1-3-7-6-9-8-2-10-5-4-12-11 | 1 165 845 |
| 6229 | 1-9-8-3-5-6-7-10-2-4-12-11 | 1 165 850 |
| 6230 | 1-8-9-6-7-3-5-10-2-4-12-11 | 1 165 853 |
| 6231 | 1-3-8-9-6-7-5-10-2-4-12-11 | 1 165 867 |
| 6232 | 1-9-6-5-10-2-4-8-3-7-12-11 | 1 165 873 |
| 6233 | 1-3-7-6-9-8-5-10-2-4-12-11 | 1 165 902 |
| 6234 | 1-9-6-7-3-5-10-2-8-4-12-11 | 1 165 928 |
| 6235 | 1-9-6-7-3-8-5-4-2-10-12-11 | 1 165 948 |
| 6236 | 1-9-6-7-3-8-2-10-5-4-12-11 | 1 166 065 |
| 6237 | 1-9-6-7-3-8-5-10-2-4-12-11 | 1 166 122 |

<-- exact answer (ring of the least length) [row 1]

<-- worse answer (ring of the greatest length) [row 6237]

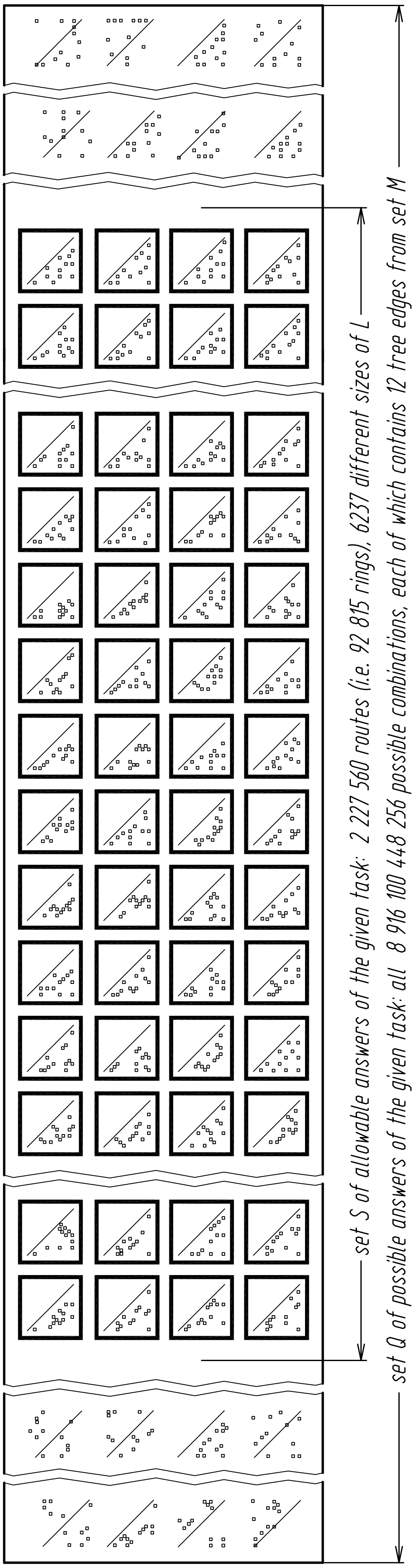

*Fig.7 Example of an individual case of the general problem TSP, i.e. one set member of the infinite set of [concrete] individual tasks about a travel salesman (lengths and positions of tree edges on the scheme of graph are shown conditionally)*

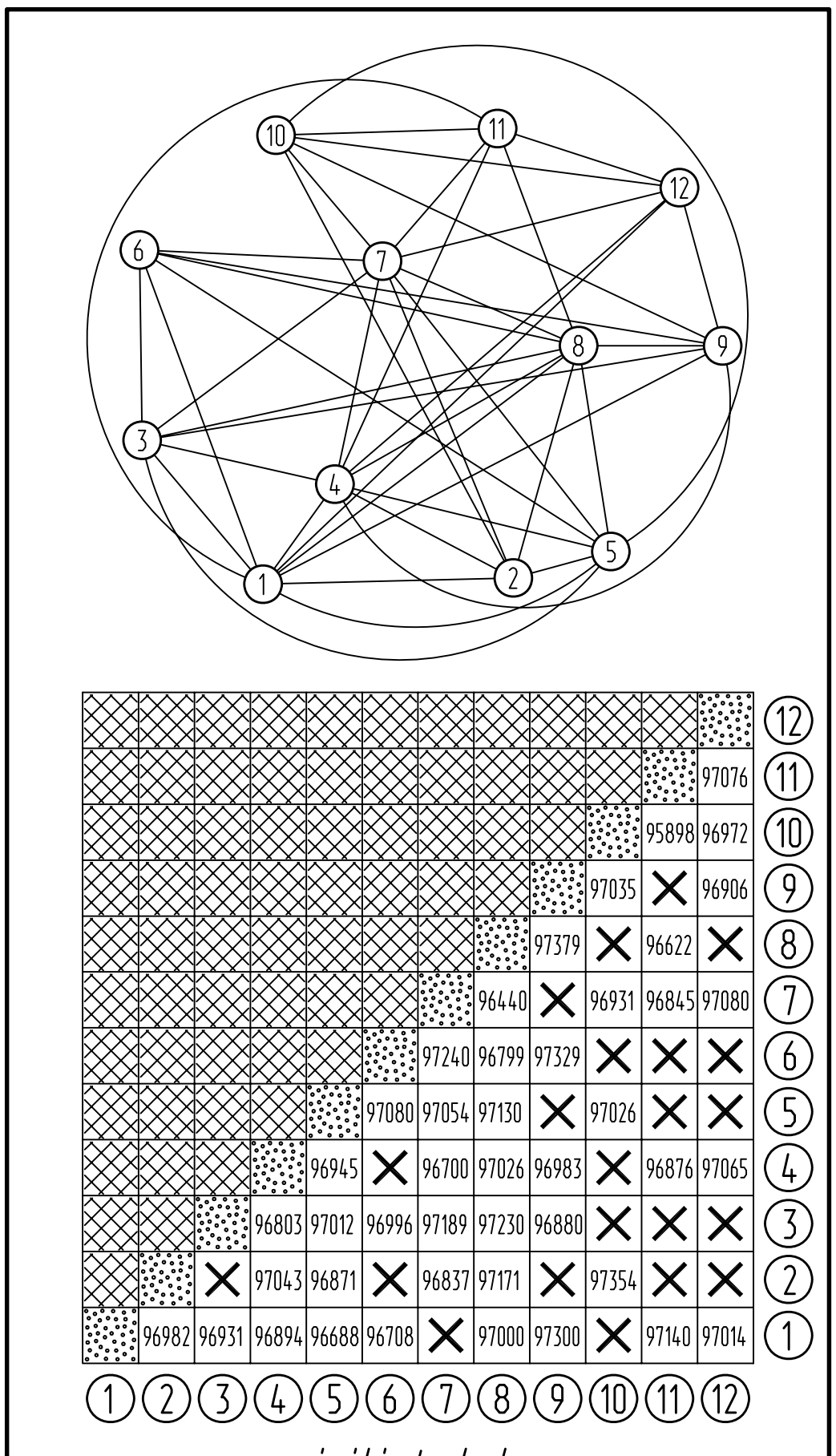

initial data

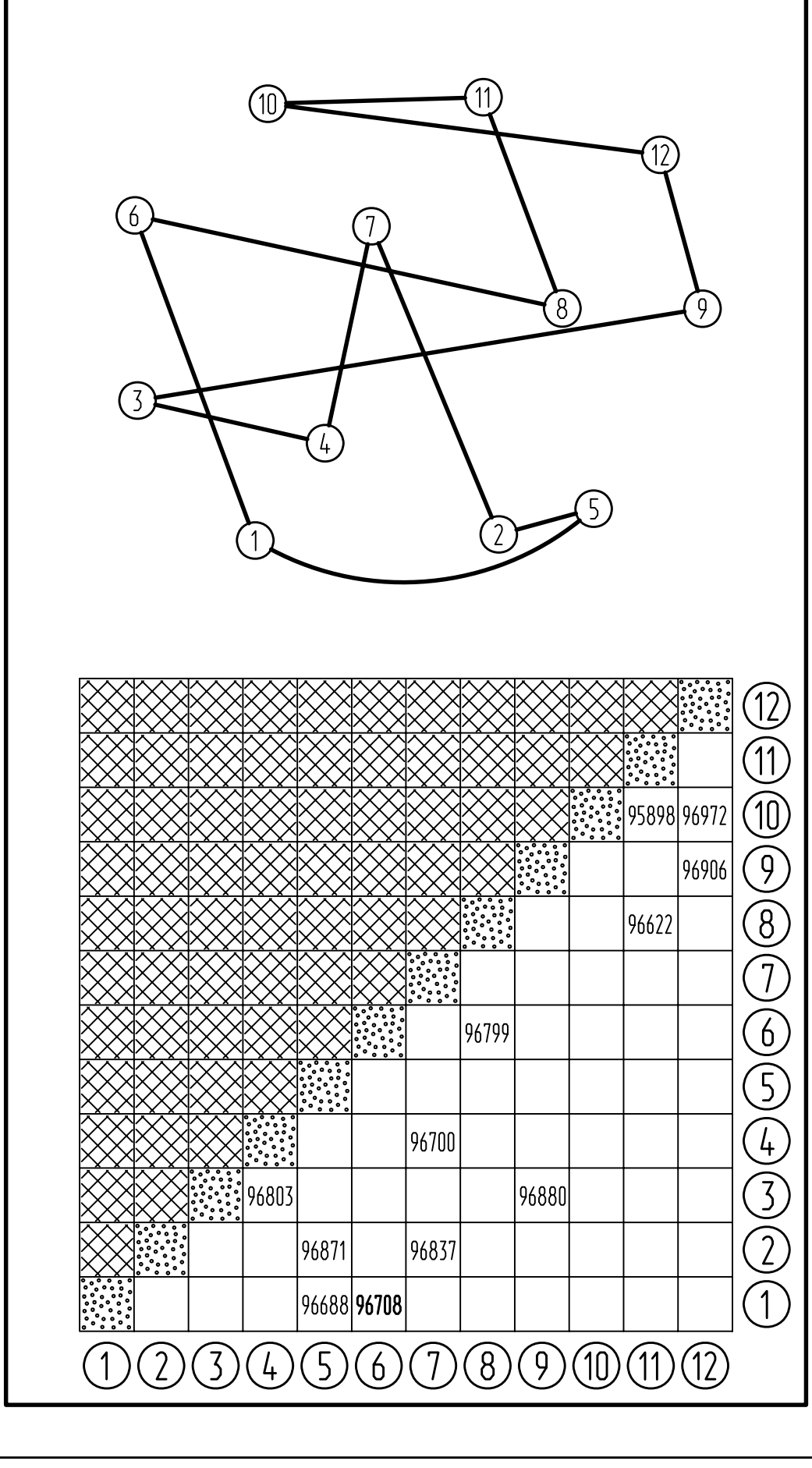

exact answer

## all allowable rings

| r. pos. of L | sequence of graph nodes | L |
|---|---|---|
| 1 | 1-5-2-7-4-3-9-12-10-11-8-6 | 1 160 684 |
| 2 | 1-5-2-4-7-8-11-10-12-9-3-6 | 1 160 724 |
| 3 | 1-5-2-4-3-9-12-10-11-7-8-6 | 1 160 853 |
| 4 | 1-5-2-7-8-11-10-9-12-4-3-6 | 1 160 869 |
| 5 | 1-4-3-9-12-10-11-8-7-2-5-6 | 1 160 911 |
| 6 | 1-2-5-4-7-8-11-10-12-9-3-6 | 1 160 920 |
| 7 | 1-5-4-2-7-8-11-10-12-9-3-6 | 1 160 935 |
| 8 | 1-5-2-7-8-11-10-12-4-9-3-6 | 1 160 960 |
| 9 | 1-4-5-2-7-8-11-10-12-9-3-6 | 1 160 969 |
| 10 | 1-5-2-4-3-9-12-11-10-7-8-6 | 1 161 043 |
| 11 | 1-2-5-3-9-12-10-11-4-7-8-6 | 1 161 044 |
| 12 | 1-2-5-4-3-9-12-10-11-7-8-6 | 1 161 049 |
| 13 | 1-3-4-9-12-10-11-8-7-2-5-6 | 1 161 051 |
| 14 | 1-4-11-10-12-9-3-6-8-7-2-5 | 1 161 057 |
| 15 | 1-6-8-7-2-5-10-11-4-3-9-12 | 1 161 058 |
| 16 | 1-5-3-9-12-10-11-4-2-7-8-6 | 1 161 059 |
| 17 | 1-3-9-12-10-11-4-5-2-7-8-6 | 1 161 063 |
| 18 | 1-6-8-11-10-5-2-7-4-3-9-12 | 1 161 064 |
| 19 | 1-5-2-4-3-9-12-10-11-8-7-6 | 1 161 071 |
| 20 | 1-2-7-8-11-10-12-9-3-4-5-6 | 1 161 073 |
| 21 | 1-6-3-4-5-2-7-8-11-10-9-12 | 1 161 075 |
| 22 | 1-2-5-10-11-12-9-3-4-7-8-6 | 1 161 089 |
| 23 | 1-4-11-10-12-9-3-5-2-7-8-6 | 1 161 093 |
| 24 | 1-2-5-10-11-8-7-4-12-9-3-6 | 1 161 094 |
| 25 | 1-5-2-7-8-4-11-10-12-9-3-6 | 1 161 098 |
| 26 | 1-4-7-2-5-3-9-12-10-11-8-6 | 1 161 099 |
| 27 | 1-5-2-7-4-8-11-10-12-9-3-6 | 1 161 104 |
| 28 | 1-5-2-8-7-4-11-10-12-9-3-6 | 1 161 106 |
| 29 | 1-5-10-11-8-7-2-4-12-9-3-6 | 1 161 109 |
| 30 | 1-5-2-7-8-11-10-12-4-3-9-6 | 1 161 113 |
| 31 | 1-5-2-10-11-12-9-3-4-7-8-6 | 1 161 123 |
| 32 | 1-5-2-10-11-8-7-4-12-9-3-6 | 1 161 128 |
| 33 | 1-4-3-9-12-11-10-5-2-7-8-6 | 1 161 138 |
| 34 | 1-4-9-12-10-11-8-7-2-5-3-6 | 1 161 139 |
| 35 | 1-5-2-7-8-6-3-4-11-10-9-12 | 1 161 163 |
| 36 | 1-6-3-9-4-5-2-7-8-11-10-12 | 1 161 166 |
| 37 | 1-2-7-4-3-9-12-10-11-8-6-5 | 1 161 167 |
| 38 | 1-5-2-7-4-3-6-8-11-10-9-12 | 1 161 169 |
| 39 | 1-2-5-7-4-3-9-12-10-11-8-6 | 1 161 195 |
| 40 | 1-3-4-12-9-10-11-8-7-2-5-6 | 1 161 196 |
| 41 | 1-2-3-4-5-6-7-8-9-10-11-12 | 1 161 199 |
| 42 | 1-6-8-7-2-5-3-4-11-10-9-12 | 1 161 203 |
| 43 | 1-2-4-7-8-11-10-12-9-3-6-5 | 1 161 207 |
| 44 | 1-5-7-2-4-3-9-12-10-11-8-6 | 1 161 210 |
| 45 | 1-6-8-11-10-7-2-5-4-3-9-12 | 1 161 214 |
| 46 | 1-2-5-3-4-9-12-10-11-7-8-6 | 1 161 219 |
| 47 | 1-5-2-4-3-9-12-7-10-11-8-6 | 1 161 229 |
| 48 | 1-5-2-7-11-10-12-9-3-4-8-6 | 1 161 233 |
| 49 | 1-2-5-7-8-11-10-12-9-4-3-6 | 1 161 235 |
| 50 | 1-3-9-12-4-11-10-5-2-7-8-6 | 1 161 237 |
| 51 | 1-2-5-4-3-9-12-11-10-7-8-6 | 1 161 239 |
| 52 | 1-2-5-6-3-9-12-10-11-8-7-4 | 1 161 241 |
| 53 | 1-2-4-7-8-11-10-12-9-3-5-6 | 1 161 243 |
| 54 | 1-6-3-5-2-4-7-8-11-10-9-12 | 1 161 245 |
| 55 | 1-6-5-2-7-8-11-10-9-3-4-12 | 1 161 253 |
| 56 | 1-5-2-7-8-6-3-9-4-11-10-12 | 1 161 254 |
| 57 | 1-4-2-7-8-11-10-12-9-3-6-5 | 1 161 256 |
| 58 | 1-6-5-10-11-8-7-2-4-3-9-12 | 1 161 257 |
| 59 | 1-5-3-4-2-7-8-11-10-12-9-6 | 1 161 258 |
| 60 | 1-5-2-7-4-9-3-6-8-11-10-12 | 1 161 260 |
| 61 | 1-2-7-4-5-3-9-12-10-11-8-6 | 1 161 261 |
| 62 | 1-3-4-5-2-7-8-11-10-12-9-6 | 1 161 262 |
| 63 | 1-2-5-4-3-9-12-10-11-8-7-6 | 1 161 267 |
| 64 | 1-4-2-5-3-9-12-10-11-7-8-6 | 1 161 268 |
| 5982 | 1-9-8-2-10-5-6-7-3-4-12-11 | 1 165 823 |
| 5983 | 1-9-10-2-8-3-7-6-5-4-12-11 | 1 165 825 |
| 5984 | 1-9-6-5-10-2-8-3-7-12-4-11 | 1 165 840 |
| 5985 | 1-3-7-6-9-8-2-10-5-4-12-11 | 1 165 845 |
| 5986 | 1-9-8-3-5-6-7-10-2-4-12-11 | 1 165 850 |
| 5987 | 1-8-9-6-7-3-5-10-2-4-12-11 | 1 165 853 |
| 5988 | 1-3-8-9-6-7-5-10-2-4-12-11 | 1 165 867 |
| 5989 | 1-9-6-5-10-2-4-8-3-7-12-11 | 1 165 873 |
| 5990 | 1-3-7-6-9-8-5-10-2-4-12-11 | 1 165 902 |
| 5991 | 1-9-8-3-6-7-5-10-2-4-12-11 | 1 165 903 |
| 5992 | 1-9-6-7-3-5-10-2-8-4-12-11 | 1 165 928 |
| 5993 | 1-9-6-7-3-8-5-4-2-10-12-11 | 1 165 948 |
| 5994 | 1-9-6-7-3-8-2-10-5-4-12-11 | 1 166 065 |
| 5995 | 1-9-6-7-3-8-5-10-2-4-12-11 | 1 166 122 |

<-- exact answer (ring of the least length)

<-- worse answer (ring of the greatest length)

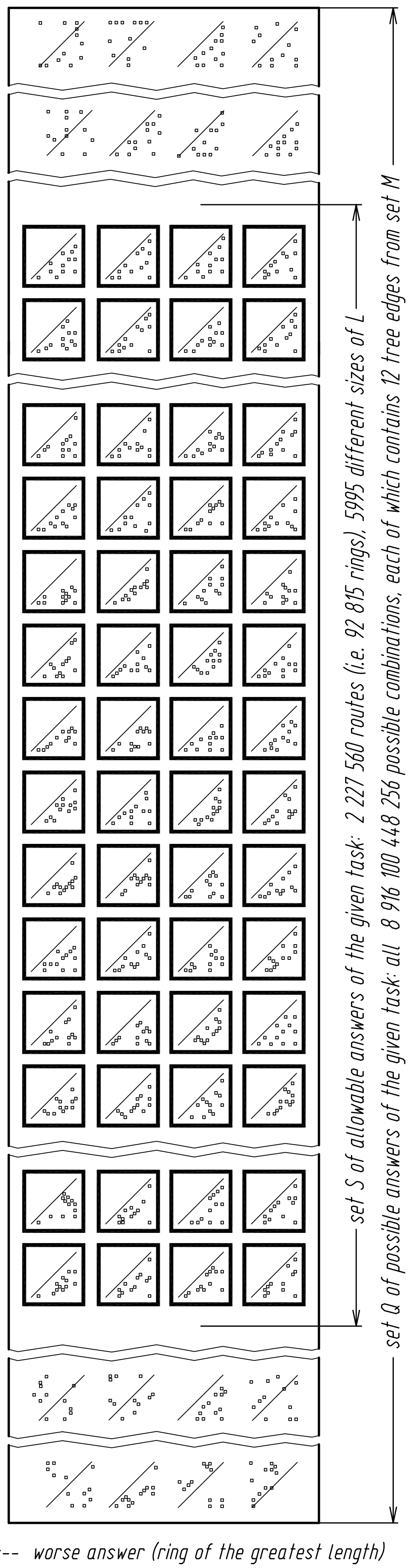

Fig.8 Other individual case of the general problem TSP. Only one difference of its initial data from the initial data of the previous example (see Fig.7): the tree edge " 3 - 6 " is longer on 2,32 %.

data are insignificant, in Fig.8 only one of edges became longer only for 2,32% (the standard accuracy, for example, in engineering practice makes 5%).

Usually for the mathematical analysis discrete and continuous tasks like EAP such little change of an independent variable gives small and analytically quite predictable (calculated) change of parameters of exact answer objectively existing in a task (change of values of required unknown). But here the exact answer (i.e. the sum of lengths of edges of the graph in the shortest of routes and that isn't less important, their sequence) significantly changed: L = 1 160 684, "1-5-2-7-4-3-9-12-10-11-8-6". Essential change of sequence (i.e. combinations) edges in the exact answer means, that the exact answer even at negligible change of an independent variable was in absolutely another and essentially unpredictable for analytical (computing) ways of the solution of the task, a point of N-dimensional space of an individual case of the general problem TSP.

The route "1-5-2-4-7-8-11-10-12-9-3-6" which in Fig.7 was the shortest, in Fig.8 only accidently for the developer and the user of algorithm appeared the owner of a rating position 2 (this route could be much farther from the exact answer). While the present exact answer earlier, in Fig.7, had more than a mediocre rating position 1340.

Comparing figuratively sets of possible and allowable answers of any individual case of TSP according to the audience on tribunes in the filled stadium and athletes-stayers, according to above-mentioned nonstrict definition of tasks of the class NP by the champion (i.e. the exact answer) it is advisable to declare the first met athlete (even if he in reality is the last of outsiders). And the business, it is very easy to check that he doesn't sit on tribunes and is dressed in sportswear. But the core of the problem of "P vs NP" here is a definition of its place in hierarchy among exponential (i.e. inaccessible in general case) a great number of other owners of sportswear. Incorrectness of the widely used informal formulation (see section 1) that in tasks like TSP it suits only for check of an allowability of answers. Such check and really is easily feasible, but it is useless from the point of view of an assessment of accuracy of the shown answer of a task.

The accuracy got by the best algorithm of the solution of the task TSP (i.e. differences of parameters of the answer received by algorithm from the appropriate parameters of the exact answer):

if it is proved, that «P = NP», than differences do not present (the received answer always represents just the exact answer);

if it is proved, that «P $\neq$ NP», than in general case the question on accuracy of the answer of the task TSP received by any algorithm except for exhaustive search, has no sense.

## 5.3. NP-complete task with additional not-quantitative requirements to the exact answer

***The task «Educational Schedule» (ESh)*** - five finite sets (student's groups, teachers, intervals of astronomical time within a week, subjects, audiences) and finite sets of wishes and the ban declared for each member of these sets and at each combination of members of these sets are set. It is necessary to make optimum from the point of view of wishes and a ban week lesson schedule at school or university. "The educational schedule" represents "not-Boolean" NP-complete task with additional not quantitative requirements. Such parameters not giving in to the full-fledged numerical description here are wishes and a ban: hours when specific teachers, specialization of teachers can teach, specializations of student's groups, incompatibility or repeatability of subjects, possibility of use of free audience, a categorical ban on existence of gaps at any student's group during the day, etc.

Here:

Allowable answers: can exist and can not exist.

Exact answer: can exist and can not exist.

In the task ESh because of its nature the developer and even the user of algorithm is free to (i.e. it is subjective) appoint criteria by which among a huge set of allowable answers it is possible to identify the required exact answer (i.e. the educational schedule which will correspond to all inconsistent ban stated by all interested parties, requirements and wishes). These criteria can be expressed in any form, even in the form of the certain quantitative coefficients which are conditionally attributed to these or those concrete values of those or independent variables. However it can happen that in a considered individual case of the task ESh there is no allowable answer from that combination of a ban, requirements and wishes which subjectively and very precipitately were declared by the developer or the user. Such situation is possible not only in the ESh.

The fact of existence of the exact answer of a considered individual case of the general problem: beforehand is not known.

Exhaustive search: if in a considered individual case among other allowable answers there is combination of parameters which satisfies to all representations of the developer or/and the user of algorithm about the exact answer, then given algorithm receives the exact answer (and also all allowable answers).

Exact polynomial (effective) algorithm: now it is not known.

Exact exponential (inefficient) algorithm which is not exhaustive search: now it is not known.

Approached polynomial (effective) algorithm: it is known.

Approached exponential (inefficient) algorithm which is not being exhaustive search: it is known.

Availability of the exact answer of the task prior to the beginning of work of algorithm (i.e. knowledge of the parameters completely determining the exact answer): parameters are unknown.

Availability of the exact answer of the task after the ending of work of algorithm:

if it is proved, that «P = NP», than parameters are known (the received answer always represents just the exact answer);

if it is proved, that «P $\neq$ NP», than in general case the exact answer of the task ESh is inaccessible.

The accuracy achievable by the best algorithm of the solution of the task ESh (i.e. differences of parameters of the answer received by algorithm from the appropriate parameters of the exact answer):

if it is proved, that «P = NP», than differences do not present (the received answer always represents just the exact answer);

if it is  proved, that «P $\neq$ NP», than, as well as for the task TSP, the question on accuracy of the answer of the task ESh received by any algorithm except for exhaustive search,  has no sense.

## Conclusions

1. Well-known informal nonstrict definition of tasks of the class NP is fair only for "Boolean" NP-complete tasks without additional requirements to their exact answer (i.e. for the tasks, similar SAT – they mean only "YES" or "NO" for an assessment of compliance obtained to requirements to which has to satisfy the exact answer of this task). Except NP-complete tasks with such “black-and-white" binary logic exist not-Boolean NP-complete tasks with additional quantitative or/and not-quantitative requirements to the exact answer;  for them this nonstrict definition is insufficiently correct theoretical and is very doubtful from the point of view of pragmatic.

2. If it is reliably proved that a solution of the problem "P vs NP" is expression "P = NP" (see [5], [7], [9], [11] etc), for TSP, ESH and other tasks of the class NP with additional requirements to the exact answer this answer can be received, as well as in other tasks of a class P, by means of polynomial algorithm.

3. If the problem "P vs NP" is successfully solved as "P $\neq$ NP" (see, for example, [6], [8], [10] etc), it will mean that for any tasks of the class NP with additional requirements to the exact answer this answer can be received only incidentally with probability, generally the inversely proportional power of a set of possible answers of a task. But even if this ever infinitely vanishing probability at the solution of any specific objective is realized, the exact answer received will be impossible to identify how exactly the exact answer is.

December, 2013.